# Collectively Driven Optical Nanoantennas

Jian Wen Choong[1], Nikita Nefedkin[2,3], and Alex Krasnok[2]

[1]Department of Electrical Engineering, The City College of New York, NY, 10031, USA

[2]Photonics Initiative, Advanced Science Research Center, City University of New York, NY 10031, USA

[3]Moscow Institute of Physics and Technology, Moscow 141700, Russia

*To whom correspondence should be addressed: akrasnok@gc.cuny.edu

**Abstract**

Optical nanoantennas, i.e., elements transforming localized light or waveguide modes into freely propagating fields and vice versa, are vital components for modern nanophotonics. Optical antennas have been demonstrated to cause the Dicke superradiance effect, i.e., collective spontaneous emission of quantum sources. However, the impact of coherent excitation on the antenna performance, such as directivity, efficiency, and Purcell effect, remains mostly unexplored. Herein, using full-wave numerical simulations backed by a quantum model, we unveil that coherent excitation allows controlling antenna multipoles, on-demand excitation of nonradiative states, enhanced directivity and improves antenna radiation efficiency. This collective excitation corresponds to the states with nonzero dipole moment in the quantum picture, where the quantum phase is well defined. The results of this work bring another degree of freedom - the collective phase of an ensemble of quantum emitters - to control optical nanoantennas and, as such, pave the way to the use of collective excitations for nanophotonic devices with superb performance. To make the discussion independent of the frequency range, we consider the all-dielectric design and use dimensionless units.

1. Introduction

Antennas are crucial for many vital wireless technologies, including communications and power transfer [1]. Being dictated by applications, a plethora of antennas in the radio and microwave frequency ranges have been invented, including microstrip antennas [2], reflector antennas [1,3], dielectric antennas [4,5], to mention just a few. More recently, the optical counterpart, the so-called nanoantenna, has also been invented for quantum optics, spectroscopy, and communications on a chip [6–12]. Plasmonic nanoantennas made of noble metals have been demonstrated to dramatically enhance light-matter interactions, a phenomenon that lies in the heart of many modern experimental techniques and applications [13–16]. Later on, all-dielectric nanoantennas have been suggested to get around the issue of material loss in metals and, as such, have found a number of prospective applications as individual elements [8,9,17–24] and as building blocks (meta-atoms) for metasurfaces [25–28].

Traditionally, nanoantennas are fed by a single quantum optical source (e.g., molecules, QDs) or by an ensemble of incoherent sources. In this scenario, the antenna-effect consists in the emission enhancement via the so-called Purcell effect, i.e., increasing the radiative decay rate of a source induced by the enlarged local density of optical states (LDOS) [29–37]. Even though the Purcell effect can lead to significant enhancement of the emitted power ($P_{rad}$) [38–40], it scales with the number of quantum sources ($N$) as $P_{rad} \propto N$, due to the incoherent nature of spontaneous emission. In 1954, Robert Dicke theoretically demonstrated [41] that in a system of $N$ excited two-level atoms, the spontaneous emission can become correlated. As a result, the entire system radiates with a dipole $d \sim N d_0$ ($d_0$ is the dipole moment of a single atom), and hence scales as $P_{rad} \propto N^2$. In the time-resolved scenario, it leads to an increase in emission rate and narrowing of the emitted pulse [42,43]. The synchronization of spontaneous emission can arise in ensembles of atoms confined in a subwavelength region of volume $< (\lambda)^3$, where $\lambda$ is the radiation wavelength. Interestingly, resonant optical nanostructures can cause correlated spontaneous emission of coupled sources [44–50]. The Dicke radiation has been predicted and observed for $N = 2$ sources [51] and ensembles of many $N \gg 1$ sources [43] in a variety of systems, including atoms [52,53], ions [51], quantum dots [43], qubits [54], and Josephson junctions [55].

It worth noting that the superradiance effect has an analogy in classic antenna theory and consists in the mutual matching of coherently excited closely arranged antennas (i.e., antenna arrays) [56] or enlarging of elastic scattering in arrays of (nano) particles [57] and on-chip photonic crystals [58]. This effect plays an essential role in Josephson-junction arrays [55] and is utilized for the emission of highly intense Cherenkov pulses [59]. In analogy with its quantum counterpart, rapid superlinear enhancement of the emission power with the number of antennas or scatterers occurs in this regime. A good discussion on the classical analogy of the Dicke radiation can be found in Refs. [47,60].

Simultaneous excitation of an antenna by several coherent sources is expected to alter its performance (e.g., directivity, efficiency, Purcell effect) by changing the multipole composition. For example, symmetric excitation of a dipole mode by two dipole sources may cause destructive interference resulting in the mode suppression. Relative coherent effects have been recently applied for perfect absorption [61–64], ultimate all-optical light manipulation [65], and enhance wireless power transfer [66,67].

In this paper, we show that coherent excitation of an antenna by two localized sources makes it feasible to *control the excitation of multipoles* and, as a result, the antenna's properties. It leads to the coherent tuning of radiated power from almost zero values (subradiance) to significantly enhanced (superradiance). Interestingly, it makes possible excitation of anapole state and turning it on/off at will. Further, we demonstrate that coherent excitation reduces dissipation losses and *improves radiation efficiency* via suppressing higher-order modes. Finally, we design an antenna operating in *superdirective and superradiance regimes simultaneously*. To make the discussion independent of the frequency range, we consider the all-dielectric design and use dimensionless units.

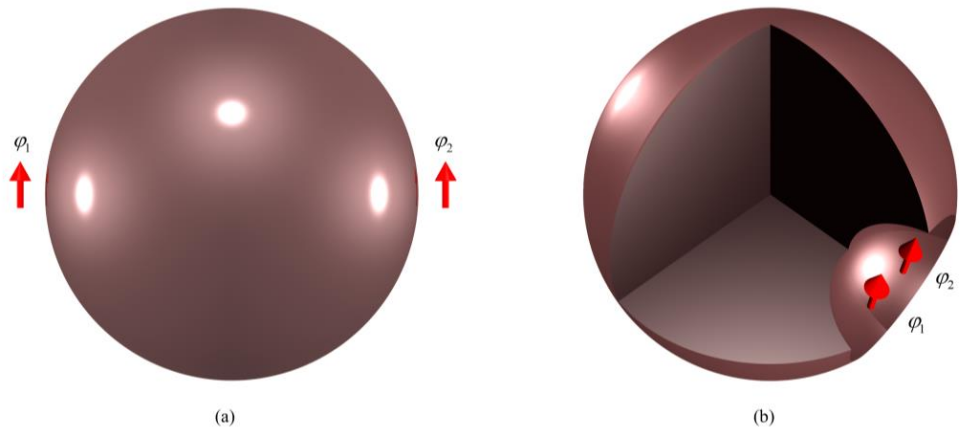

FIG. 1. Schematic representation of considered (a) dielectric antenna and (b) notched superdirective antenna coherently driven by two dipole sources.

## 2. Results and discussion

**2.1 Sub- and superradiance.**

Let us consider a high-index dielectric resonator of radius $R$ and refractive index $n=4$ exited by two coherent dipole sources, Fig. 1(a). The chosen refractive index corresponds to traditionally used materials in optics and microwaves [68–70]. Such a dielectric resonator supports Mie modes of a different order [see Fig. 7(a) in Appendix A]. The first mode, magnetic dipole (md), is excited when the radius satisfies $R \approx \lambda/2n$ (for $n=4$, it gives $\lambda/R \approx 8$) [25,28]. Higher-order modes: electric dipole (ed), magnetic quadrupole (mq), etc., appear at shorter wavelengths.

First, it is illustrative to consider a scenario of excitation by two oppositely directed plane waves. For simplicity, we assume the resonator supports only ed and md modes. Upon excitation, due to the linearity, the absolute values of Mie scattering dipole electric and magnetic amplitudes can be expressed as $|a_1 + a_1 e^{i\varphi}|$ and $|b_1 - b_1 e^{i\varphi}|$, respectively [see Fig. 7(b) and (c) in Appendix A]. The different signs stem from the pseudo-vector character of the md. Thus, this scenario illustrates an ability to coherent control of the antenna modes by two-wave excitation with the relative phase ($\varphi$).

Next, we assume the resonator is excited by two dipoles of equal amplitude ($P_{dy}$, $|P_{dy}|=1$) but different phases ($\varphi_1$ and $\varphi_2$), and polarized along the y-axis, as shown in Fig. 1(a). Below we introduce the *states with nonzero dipole moment* [71] as an initial state, where the phase is well defined and, hence, provide the fully quantum description. The resonator supports md and ed moments with the magnetic ($\alpha_p^m$) and electric ($\alpha_p^e$) polarizabilities and corresponding magnetic ($M_p$) and electric ($P_p$) dipole moments. This system can be rigorously described by the discrete dipole approximation (DDA) approach [22,72] (see Appendix B for details), which yields the following nonzero components of the electric and magnetic dipole moments

$$P_{py} = \alpha_p^e A_{pd} P_{dy} (e^{j\varphi_1} + e^{j\varphi_2}) \tag{1}$$

$$M_{pz} = \alpha_p^m \sqrt{\frac{\varepsilon_0}{\mu_0}} D_{pd} P_{dy} (e^{j\varphi_1} - e^{j\varphi_2}) \tag{2}$$

where $\varepsilon_0$ and $\mu_0$ are the permittivity and permeability of free space; $A_{pd}$ and $D_{pd}$ are frequency depending coefficients defined in Appendix B. When the dipole sources radiate in-phase ($\varphi_1 - \varphi_2 = 0$), the magnetic dipole moment of the resonator vanishes, $M_{pz} = 0$, while the electric dipole moment is twice larger than in the case of single-source excitation ($|P_{py}| = 2|\alpha_p^e A_{pd}|$). In the case of $|\varphi_1 - \varphi_2|=180$ deg, the resonator possesses zero electric dipole moment and enhanced magnetic dipole ($|M_{pz}|=2\sqrt{\varepsilon_0/\mu_0}\,|\alpha_p^m D_{pd}|$).

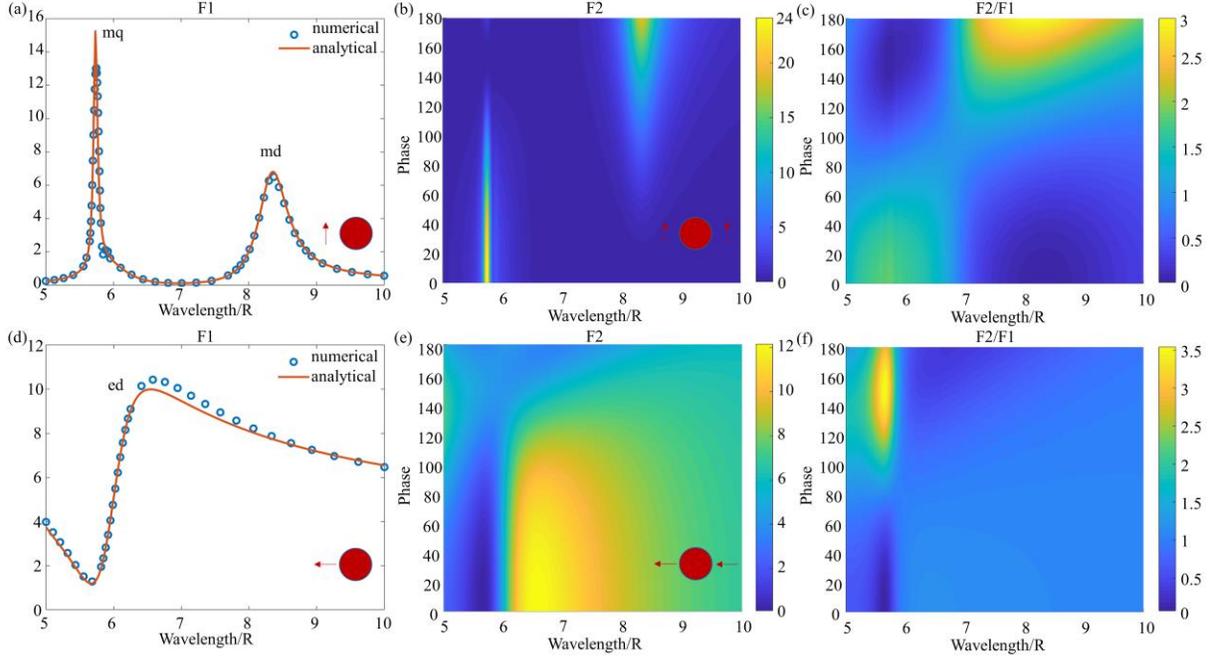

FIG. 2. Purcell factor of a single dipole source ($F_1$) with (a) tangential (TD) and (d) longitudinal (LD) orientation vs. wavelength normalized to the radius ($\lambda/R$) calculated both numerically and analytically. (b), (e) Collective radiation enhancement of two dipoles ($F_2$) with (b) TD and (e) LD orientation vs. $\lambda/R$ and phase difference of dipoles ($\varphi_d = |\varphi_1 - \varphi_2|$). (c), (f) Normalized collective radiation enhancement ($F_2/F_1$) for (c) TD and (f) LD orientation vs. $\lambda/R$ and $\varphi_d$. The actual radius is R = 60 nm; the dipole sources of $l_d$ = 20 nm are located at a distance of 10 nm from the surface.

Next, to explore how this coherent excitation affects the antenna properties, we start with calculating the Purcell effect. Firstly, we calculate the Purcell factor ($F_1$) for the single dipole source for both tangential (TD) and longitudinal (LD) orientation, Figs. 2(a,d). For numerical calculation, we use CST Microwave Studio and the input-impedance approach reported in Ref. [36]. Here we consider first a lossless system, whereas the effect of loss is discussed in what follows. For a lossless system, the Purcell factor coincides with the ratio [36]: $F_1 = P_{rad}/P_{0,rad}$, where $P_{rad}$ is the radiated power for the presence of resonator and $P_{0,rad}$ is for free space.

The results of the numerical calculation of the Purcell factor of a single dipole ($F_1$) for tangential (TD) and longitudinal (LD) orientation are presented in Figs. 2(a,d) by dots. We also use the Green's function approach to verify our numerical results (solid red curves) [73,74]. For TD polarization, we observe two resonant modes, which are magnetic dipole (md) and magnetic quadrupole (mq) at $\lambda/R = 8.37$ and $\lambda/R = 5.75$, respectively, Fig. 2(a). The ed moment is not excited in TD polarization due to its zero overlap with the source, while effectively excited at $\lambda/R = 6.58$ for LD polarization, Fig 2(d). These excited resonant modes lead to a dramatic increase in power radiation and Purcell factor for both TD and LD polarisations. These results coincide with the analytical ones (red curve) and previous works [25,75].

The Purcell effect can be further increased by introducing another emitter, Fig. 1(a). To describe this case, we introduce the *collective radiation enhancement* $F_2$ factor

$$F_2(\varphi_d) = \frac{P_{\text{rad}}(\varphi_d)}{P_{0,\text{rad}}(\varphi_d)} \qquad (3)$$

where $P_{\text{rad}}(\varphi_d)$ [$P_{0,\text{rad}}(\varphi_d)$] is the collective radiated power (for the phase difference $\varphi_d$) with [without] antenna. The results of the calculation of this quantity versus $\lambda/R$ and phase difference ($\varphi_d$) are presented in Figs. 2(b,e). We observe a strong enhancement of $F_2$ up to 24 at mq ($\lambda/R = 5.75$) and 20 at md ($\lambda/R = 8.37$) modes, which is ~85% and ~300% increase compared to $F_1$ for TD, Fig. 2(b). This enhancement happens at certain relative phases. On the other hand, $F_2$ reaches 12 for ed for LD orientation, which is ~20% enhancement, Fig. 2(e).

The ratio $F_2/F_1$ gives the normalized collective radiation enhancement. If $F_2/F_1 > 1$ ($F_2/F_1 < 1$), the antenna boosts (suppresses) the collective emission. The results presented in Figs. 2(c,f) show that the antenna boosts collective emission at certain phases. Remarkably, at $\lambda/R = 5.68$ where $F_1$ is minimal, adding the second source with $\varphi_d = 150$ deg can increase the radiated power by 3.5 [Fig. 2(f)] associated with a nonradiative anapole state as discussed in what follows.

Besides collective radiation enhancement, manipulating the phase difference can also achieve the subradiance effect, which technically "turns off" the antenna. Indeed, $F_2$ can be lowered down to ~0 (at $\varphi_d = 155$ deg). Also, while $F_2$ features a 300% boost at $\lambda/R = 8.37$ for $\varphi_d = 180$ deg, zero power radiation can be achieved by altering the relative phase to be the same. A similar effect is also noticed at $\lambda/R = 5.68$ for LD. These results suggest that one can tune the Purcell factor and emitted power by adjusting the relative phase of emitters.

**2.2 Quantum justification.**

Next, we demonstrate that the observed effects remain fair in the quantum description. Due to the generality of the system under consideration, it can be treated in the field of quantum optics, and the effects obtained can also be found in the quantum case. Consider the quantum analog of the system: two quantum emitters and a dipole mode of the antenna. We assume the emitters to be two-level atoms with the same transition frequency $\omega_A$, and we suppose only the electric dipole mode of the antenna with frequency $\omega_M$. The lowering, $\hat{\sigma}_1 = \hat{\sigma} \otimes \hat{I}$ and $\hat{\sigma}_2 = \hat{I} \otimes \hat{\sigma}$, and raising, $\hat{\sigma}_1^+ = \hat{\sigma}^+ \otimes \hat{I}$ and $\hat{\sigma}_2^+ = \hat{I} \otimes \hat{\sigma}^+$, operators describe the relaxation and excitation of the 1 or 2 atom, where $\hat{I}$ is the identity matrix of size $2 \times 2$, and $\hat{\sigma} = |g\rangle\langle e|$, $\hat{\sigma}^+ = |e\rangle\langle g|$ are transition operators from the excited $|e\rangle$ state to the ground $|g\rangle$ state and vice versa. Operators $\hat{a}$ and $\hat{a}^+$ are annihilation and creation operators of a quantum in the antenna mode. The interaction Hamiltonian of the emitters with the resonator mode has the Jaynes-Cummings type in the rotating wave approximation. The system also interacts with the electromagnetic modes of free space. The resulting Hamiltonian of the entire system has the form:

$$\begin{aligned}
\hat{H} &= \hat{H}_S + \hat{H}_R + \hat{H}_{SR} \\
\hat{H}_S &= \sum_{i=1}^{2} \hbar \frac{\omega_A}{2} \hat{\sigma}_z^i + \hbar \omega_M \hat{a}^+ \hat{a} + \Omega \sum_{i=1}^{2} \left( \hat{a}^+ \hat{\sigma}_i + \hat{\sigma}_i^+ \hat{a} \right) \\
\hat{H}_R &= \sum_k \hat{a}_k^+ \hat{a}_k \\
\hat{H}_{SR} &= \sum_{k,i} \kappa_k (\hat{\sigma}_i + \hat{\sigma}_i^+)(\hat{a}_k^+ + \hat{a}_k) + \sum_k \chi_k (\hat{a}^+ + \hat{a})(\hat{a}_k^+ + \hat{a}_k)
\end{aligned} \quad (4)$$

where $\Omega_i = -\mathbf{d}_i \mathbf{E}_M / \hbar$ is the coupling constant of the emitters and the mode. $\hat{a}_k$ and $\hat{a}_k^+$ are annihilation and creation operators of a photon in a mode with the frequency $\omega_k$. Constants $\kappa_k$ and $\chi_k$ are the interaction constants between the emitters and free-space modes and the antenna mode and free-space modes. The dynamics of the system can be described by the master equation in the Lindblad form [76]. Using the Hamiltonian (4), we apply the Born-Markov approximation [77] and obtain the final master equation:

$$\frac{\partial}{\partial t}\rho = -\frac{i}{\hbar}\left[\hat{H}_S, \rho\right] + \frac{\gamma}{2}\left(2\hat{J}^-\rho\hat{J}^+ - \hat{J}^+\hat{J}^-\rho - \rho\hat{J}^+\hat{J}^-\right) + \frac{\gamma_a}{2}\left(2\hat{a}\rho\hat{a}^+ - \hat{a}^+\hat{a}\rho - \rho\hat{a}^+\hat{a}\right) \quad (5)$$

where we introduce the collective operators of dipole moment $\hat{J}^- = \sum_i \hat{\sigma}_i$ and $\hat{J}^+ = \sum_i \hat{\sigma}_i^+$; $\gamma$ is the spontaneous emission rate of atoms into free-space modes, $\gamma_a$ is the antenna mode's decay rate. Note that we consider weak interaction between subsystems.

The emitters are initially excited by the pulse pumping into some states. The phase of the emitter, in this case, can be found with the use of the phase operator. Following [78,79] we introduce the Hermitian phase operator $\hat{\varphi}$ for $M$-level system

$$\hat{\varphi} = \left(\varphi + \pi\frac{M-1}{M}\right)\hat{I}_M + \frac{2\pi}{M}\sum_{m\neq m'}^{M-1}\frac{\exp[i(m-m')\varphi_0]}{\exp[2\pi i(m-m')/M] - 1}|m\rangle\langle m'| \quad (6)$$

, where $\hat{I}_M = \sum_{m=0}^{M-1}|m\rangle\langle m|$ and $\varphi_0$ is a reference phase (we put it as $\varphi_0 = 0$). The eigenvalues of this operator lie in the range $[\varphi_0, \varphi_0 + 2\pi)$. The detailed description of the phase operator properties can be found in Refs. [80,81].

For the two-level atom, the phase operator takes a simple form:

$$\hat{\varphi} = \begin{pmatrix} \pi/2 & -\pi/2 \\ -\pi/2 & \pi/2 \end{pmatrix} \quad (7)$$

As an illustration, one can easily obtain that the phase difference for collective Dicke state [41] of $N$ emitters $|N,n\rangle = 1/\sqrt{C_N^n}\sum_{\text{permutations}}|\underbrace{e,\cdots,e}_{n},\underbrace{g,\cdots,g}_{N-n}\rangle$ (if we define $\hat{\varphi}_N^i = \underbrace{\hat{I}\otimes\cdots\otimes}_{i-1}\hat{\varphi}\underbrace{\otimes\cdots\otimes\hat{I}}_{N-i}$) is always $\langle N,n|\hat{\varphi}_i^N - \hat{\varphi}_j^N|N,n\rangle = 0$, i.e. for the system in this state, emitters are indistinguishable and each of them has the same phase. Moreover, the collective dipole moment for this kind of collective states is always zero [78]. Therefore, we should introduce the states with nonzero dipole moment [71] as an initial state, for which we can define the initial phase. The density matrix, in this case, is a direct product of density matrices of atom 1 and atom 2 $\rho = \rho_1 \otimes \rho_2$, where

$$\rho_i = \begin{pmatrix} \beta_i & \alpha_i \exp(i\varphi_i) \\ \alpha_i \exp(-i\varphi_i) & 1 - \beta_i \end{pmatrix}, \quad (8)$$

$\beta_i$, $\alpha_i$ are real numbers, which are connected by a relation $\alpha_i^2 < \beta_i(1-\beta_i)$ in order to satisfy the positive definiteness of the density matrix, and $\varphi_i$ is a "classical" phase. Using this initial density matrix, we see that the average value of the dipole moment $\langle \hat{\sigma}_i \rangle = \text{Tr}(\hat{\sigma}_i \rho) = \alpha_i \exp(i\varphi_i)$ is nonzero.

Let us find the connection between the phase of the average value of the dipole moment $\varphi_i$ and the average value of the phase operator (7), $\langle \hat{\varphi}_i \rangle$, for the states with nonzero dipole moment (8). We apply the phase operator to the state (8):

$$\langle \hat{\varphi}_i \rangle = \text{Tr}\left( \begin{pmatrix} \beta_i & \alpha_i \exp(i\varphi_i) \\ \alpha_i \exp(-i\varphi_i) & 1-\beta_i \end{pmatrix} \begin{pmatrix} \pi/2 & -\pi/2 \\ -\pi/2 & \pi/2 \end{pmatrix} \right) = \frac{\pi}{2} - \pi \alpha_i \cos\varphi_i \qquad (9)$$

After this, one can obtain the average value of the difference of phase operators of two emitters

$$\langle \hat{\varphi}_i - \hat{\varphi}_j \rangle = -\pi \left( \alpha_i \cos\varphi_i - \alpha_j \cos\varphi_j \right) \qquad (10)$$

Thus, we have the *relation between the average values of phase operator and the phase of emitter's dipole moment and can consider the dependence of radiation enhancement on the phase difference between two quantum emitters' dipole moments*.

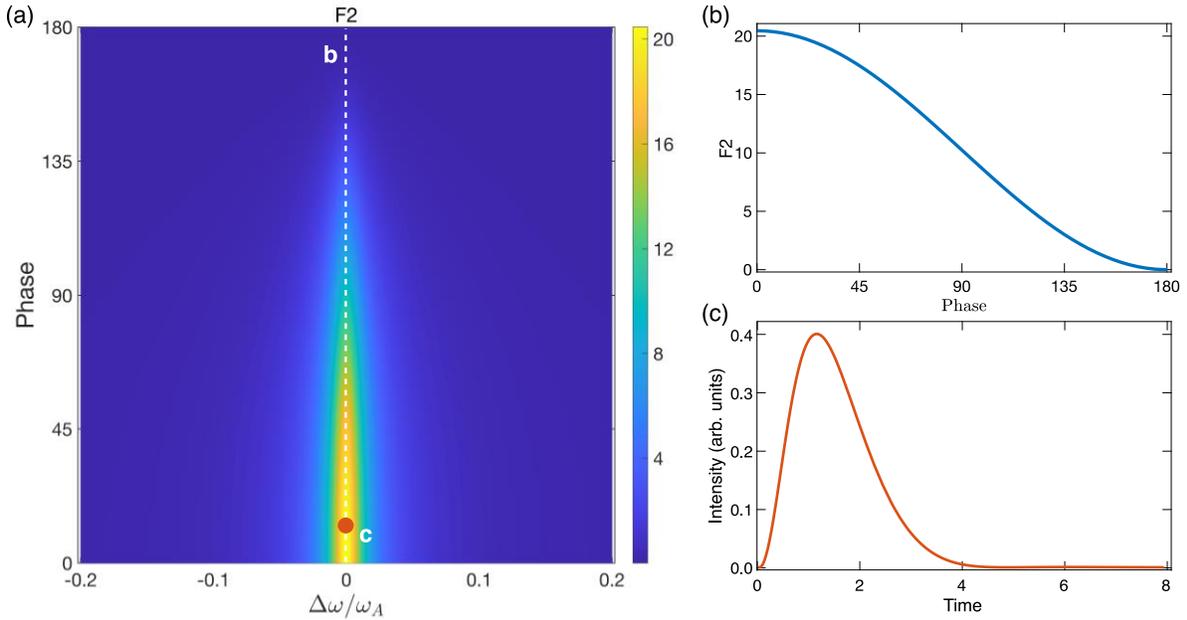

FIG. 3. (a) Collective radiation enhancement of two quantum emitters with TD orientation in dependence on the detuning between the mode and emitters ($\Delta\omega = \omega_C - \omega_A$) and the phase

difference between emitters $\Delta\varphi$. (b) Profile of $F_2$ along white dashed line. (c) Profile of antenna radiation intensity vs time at orange dot mark.

Fig. 3(a) shows the radiation enhancement $F_2$ [see Eq. (3)] in the plane of the frequency detuning between emitters and the resonator mode and the initial phase difference between emitters' dipole moments $\Delta\varphi$ [see Eq. (8)]. One can see the behavior similar to Fig. 2(b). When the frequencies of the emitters and the resonator mode are equal, $F_2$ takes the maximum values, and if the emitters' dipole moments are in phase, $F_2$ reaches $\sim 20$, which is in good agreement with the classical approach. Fig. 3(c) demonstrates the typical superradiant profile of antenna radiation intensity at resonance conditions and when the emitters' dipole moments are nearly in phase. In the opposite case, when the emitters are out of phase ($\Delta\varphi = |\varphi_1 - \varphi_2| = \pi$), the antenna radiation is significantly suppressed.

Thus, we have shown that in the quantum optical system as well as in a classical one, there is a good opportunity to control the output radiation of the antenna by tuning the initial phases of emitters. The superradiance appearing in this system leads to significant collective radiation enhancement if the initial phase difference is close to zero, and the subradiance effect emerges when the $\Delta\varphi = \pi$.

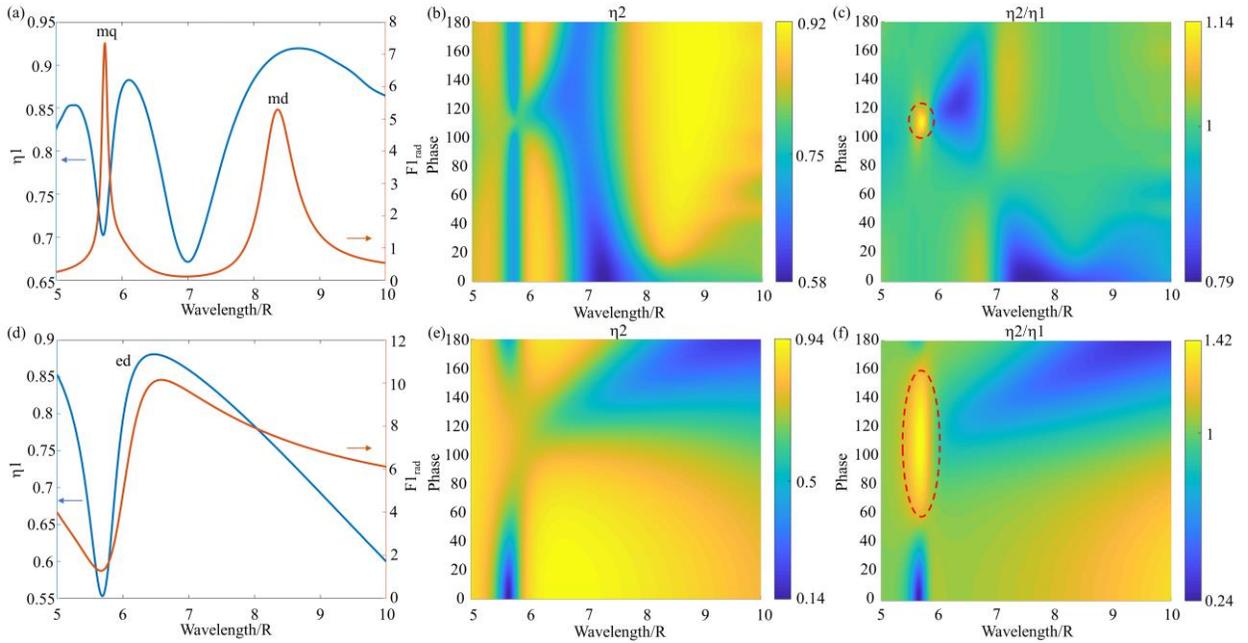

FIG. 4. Radiation efficiency and radiative Purcell factor of the antenna consisting of dipole and resonator with radius R and refractive index, $n=\sqrt{16+0.1i}$, depending on $\lambda/R$ and $\varphi_d$. (a) and (d) show both radiated efficiency (blue curve) and radiative Purcell factor (red curve) of one dipole for TD and LD orientation, respectively, depending on $\lambda/R$. (b), (e) Radiation efficiency of two dipoles ($\eta_2$) with TD and LD orientation respectively depending on $\lambda/R$ and $\varphi_d$. Normalized radiation efficiency ($\eta_2/\eta_1$) for (c) TD and (f) LD orientation depending on $\lambda/R$ and $\varphi_d$. The value of $\eta_2/\eta_1$ has the maximum at wavelength/R of 5.7 with a ~110 deg phase difference for both TD and LD orientation. The red dashed circle shows the region where the maximum $\eta_2/\eta_1$ is achieved. The dipole sources are located at the distance of 10 nm from the particle surface.

## 2.3 Efficiency boosting.

Another essential quantity characterizing any antenna is its radiation efficiency ($\eta_1$), defined as $\eta_1 = P_{rad}/P_{tot}$, where $P_{tot}$ is the total delivered power to the system. Note that although this definition is fair for both microwave and optics, in quantum optics, another definition (also called quantum efficiency) through the number of radiated photons ($N_{rad}$) is standard: $\eta_1 = N_{rad}/N_{tot}$, where $N_{tot}$ is the total number of quasiparticles (electrons, excitons). The radiation efficiency can be increased via the Purcell effect [79] by speeding up the radiative decay rate and reducing the nonradiative decays. However, in reality, the emission of a dipole source located near to a resonator surface gets dissipated due to the quenching effect, i.e., strong dissipation through excitation of nonresonant higher-order modes [82–84].

Here, we show that *collective excitation by two sources can boost radiation efficiency* via the weakening of other modes. To this end, we introduce a realistic imaginary part to the resonator refractive index, $n=\sqrt{16+0.1i}$. We calculate the radiated efficiency of one dipole ($\eta_1$) for TD and LD orientations, shown in Figs. 4(a,d) by blue curves. The radiation efficiency at the md resonance of TD orientation is ~0.7, and at the ed resonance of LD orientation is ~0.85. Consequently, the antenna dissipates a significant amount of power before radiating, even though it has a high Purcell factor, Figs. 4(a,d), red curves.

We define the *collective radiation efficiency*

$$\eta_2(\varphi_d) = \frac{P_{rad}(\varphi_d)}{P_{tot}(\varphi_d)} \tag{11}$$

where $P_{tot}(\varphi_d)$ is the phase-dependent total delivered power. The results of the numerical calculation of this value are presented in Figs. 4(b,e). To compare these results with $\eta_1$, we take their ratio, Figs. 4(c,f). We see that the presence of the second source increases the collective radiation efficiency ($\eta_2/\eta_1 > 1$) for both TD and LD orientations at certain relative phases. We observe the most significant effect at $\lambda/R = 5.7$ for both TD and LD. The efficiency of TD at mq is increased by 14% at the relative phase of 110 deg, as shown in Fig. 4(c). On the other hand, at $\lambda/R = 5.7$ of LD, the antenna that originally had a low efficiency of 55%, Fig. 4(d), has become much more (42%) efficient in the collective excitation scenario, Fig. 4(f). These results convincingly show that the collective coherent excitation can significantly boost the radiation efficiency of an antenna.

Let us consider the LD polarized excitation at $\lambda/R = 5.7$ in more detail. Figs. 5(a,b) show the collective radiation enhancement ($F_2$) and radiation efficiency ($\eta_2$) for the relative phase of 0 deg (blue curves) and 110 deg (red curves). We observe that both characteristics get increased at 110 degrees. Hence, the antenna at this wavelength radiates more and does it in a much more efficient way. To elucidate it, we show the electric field distribution at $\lambda/R = 5.7$ for both phases, Fig. 5(c). The vector E-field distribution at 0 deg corresponds to the anapole state [85] with enhanced linear distribution corresponding to the Cartesian dipole and two loops, corresponding to the toroidal moment [85,86] (see Appendix A). This anapole state is associated with increased E-field in the center, accompanied by large dissipation losses, Fig. 5(d).

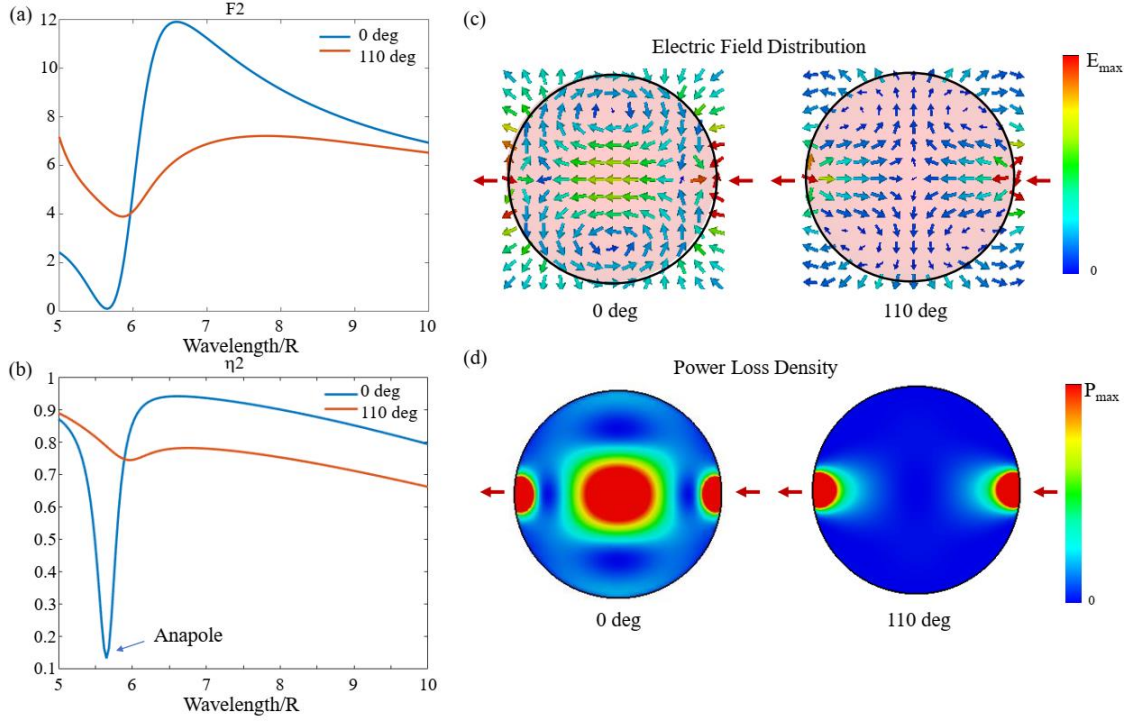

FIG. 5. (a), (b) Dependences of the collective radiation enhancement ($F_2$) and radiation efficiency ($\eta_2$) on wavelength for the relative phase of 0 deg (blue curves) and 110 deg (red curves) for the longitudinal (LD) excitation. (c), (d) E-field and power loss density in the resonator for the zero phase difference (left column) and 110 deg (right column) at $\lambda/R=5.7$. The dipole sources are located at the distance of 10 nm from the particle surface.

Figs. 5(c,d), right column, show that at the relative phase of 110 deg, the coherent excitation leads to *"turning off" the anapole state* with suppression of the E-field in the resonator center with the corresponding suppression of the power loss density and gain in both collective radiation enhancement ($F_2$) and mutual radiation efficiency ($\eta_2$) at $\lambda/R=5.7$.

**2.4 Superdirectivity.**

The above analysis shows that the coherent excitation of an antenna by two sources leads to the enhancement of radiated power and radiation efficiency. The reason behind these effects is the changing of excited multipoles. In this section, we demonstrate that the same approach provides a powerful tool for directivity engineering. To this end, we utilize the recently reported design for an all-dielectric superdirective notched antenna [20,87], Fig. 6(a). Superdirective antennas are subwavelength antennas with the directivity much larger than that of a short dipole antenna, $D_{max} = 1.5$ [20,87–93]. Superdirectivity regime relies on rapidly spatially oscillating currents and high-order multipoles in a subwavelength area [87,88]. If the multipoles are excited with certain phases and amplitudes, their far-fields interfere, forming a spatially narrow radiation beam. Here we demonstrate that the excitation of a superdirective antenna by two coherent sources allows ultimate control of excited multipoles and radiation patterns.

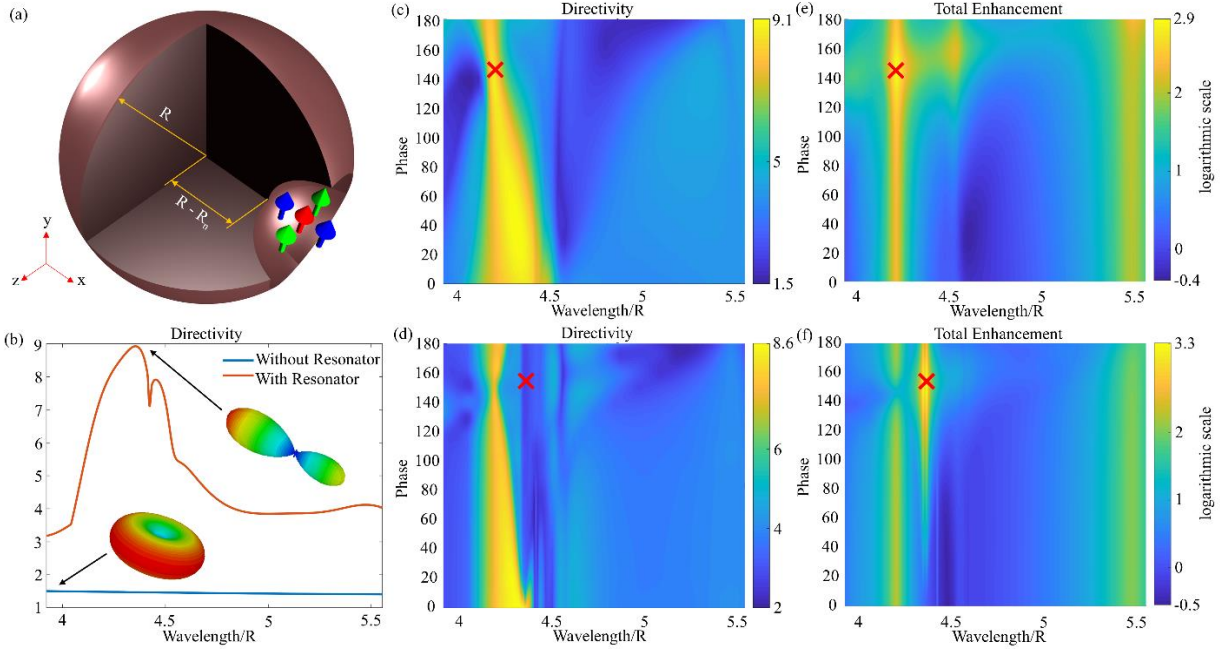

FIG. 6. (a) Schematic representation of the notched antenna comprising two coherent dipoles of different configurations. (b) Directivity of the single dipole [red arrow in (a)] with and without resonator depending on $\lambda/R$. Insets show the corresponding radiation patterns. (c), (d) Directivity of two dipoles placing along the (c) x-axis [blue arrows in (a)] and (d) z-axis [green arrows in (a)] in the notch. (e), (f) Total enhancement, $\Sigma = D_{max} \cdot F_2$, in logarithmic scale for two dipoles placing along the (e) x-axis and (f) z-axis in the notch of the resonator. The red crosses show the points of maximum total enhancement that results in superradiative and superdirective regimes simultaneously. The radius of the antenna and notch are R = 90 nm and $R_n$ = 40 nm. The notch's center is exactly on the surface of the resonator, and the midpoint between the dipoles is 20 nm away from the surface.

Following the antenna textbooks, we define the directivity as $D_{max} = 4\pi P_{max}(\theta,\varphi)/P_{rad}$, where $(\theta,\varphi)$ are the angular coordinates of the spherical coordinate system, and $P_{max}$ is the power in the direction of the main lobe [1]. This value is normalized so that the isotropic point source has $D_{max}=1$ and the dipole source has $D_{max}=1.5$. Fig. 6(b) presents the directivity of the notched antenna for single dipole excitation [Fig. 6(a), red arrow]. We observe the maximum directivity $D_{max}=9$ at $\lambda/R=4.35$ (red curve), which is much higher than that in free space (~1.5, blue curve). The insets demonstrate the corresponding radiation patterns.

The presence of the second dipole source allows coherent tuning of this superdirective antenna. For the arrangement of the sources along the x-axis [Fig. 6(a), blue arrows], we observe the preservation of the superdirectivity regime with the maximum directivity (~9.1), which changes with the relative phase, Fig. 6(c). To estimate the overall enhancement of the antenna performance, we take the product of directivity and collective radiation enhancement $\Sigma = D_{max} \cdot F_2$ (see Fig. 8 in Appendix C for details). For this x-axis orientation, the enhancement by a factor of 880 (2.9 in logarithmic scale) is achieved, Fig. 6(e). At this point, the antenna possesses both superdirectivity and superradiation effects with $F_2=100$ and $D_{max}=8.5$. When the sources are placed along the z-axis [Fig. 6(a), green arrows], a higher maximum total enhancement of 2160 (3.3 in logarithmic scale) is observed for z-axis orientation despite smaller directivity (~2.7) in this case. Thus, for the x-axis arranged coherent sources, the antenna can operate in the superradiative ($F_2/F_1>1$) and superdirective ($D_{max} \gg 1$, $\lambda/R>1$) regimes simultaneously.

## 3. Conclusions

In this paper, we have explored how the coherent excitation by several sources can affect the antenna performance. We have shown that coherent excitation of an antenna by two sources makes it feasible to control the excitation of multipoles and, as a result, its electromagnetic properties. It leads to the coherent tuning of radiated power from almost zero values (subradiance) to significantly enhanced (superradiance). We have explored that this approach reduces the quenching effect and strengthens the radiation efficiency at some specific phases via coherent cancelation of higher-order modes. The approach allows excitation of the nonradiative field configuration, anapole state, and turning it on/off at our will. In the quantum system, we have shown that this collective excitation corresponds to the states with nonzero dipole moment, where the quantum phase is well defined. We have also demonstrated that utilizing this approach allows designing an antenna operating in superdirective and superradiance regimes simultaneously with the total enhancement factor over $2\cdot 10^3$. We believe that the findings reported in this study will found applications in coherently driven antennas, active and quantum nanophotonics. Due to the dipole approximation and peculiarities of Mie resonances, the spherical antennas can be replaced by antennas of cylindrical shape [28]. While spherical antennas are feasible for analytical investigation, particles of cylindrical shape can be realized with well-established lithography and chemical etching methods.

There are several systems, which allow well control of the initial states. First, one can use the cold atoms or ions in optical traps [94–96]. The fine control of the initial state can be obtained by changing both the potential wells' configuration and the external fields. Second, the superconducting qubits platform provides a great opportunity to tune the initial states of qubits very precisely [97]. Also, among the adiabatic techniques to control the quantum states [98], which rely on the fact that the system remains in its state despite the slow changes to the system, there is a superadiabatic transitionless driving technique [99]. This method allows a fast and robust coherent quantum control, which was experimentally demonstrated in the NV centers and can be implemented in other solid-state systems.

The states with a certain phase difference are formed in a wide class of quantum emitters, including QDs, NV centers, atoms, and molecules due to interaction with the same mode of a cavity or antenna. Even though this phase difference is barely predictable, our work suggests a variety of effects that this phase would cause.

Lastly, we present our results in dimensionless units, so they are applicable to antennas in microwaves, where the control of the relative phase of the elementary emitters is a rather established technique.

**Acknowledgments**

The authors thank the CUNY Summer Undergraduate Research Program and Alfred P. Sloan Foundation for the support.

**Conflict of Interest**

The authors declare no conflict of interest.

**Appendix A: Multipoles and Coherent tuning of a dielectric nanoparticle scattering**

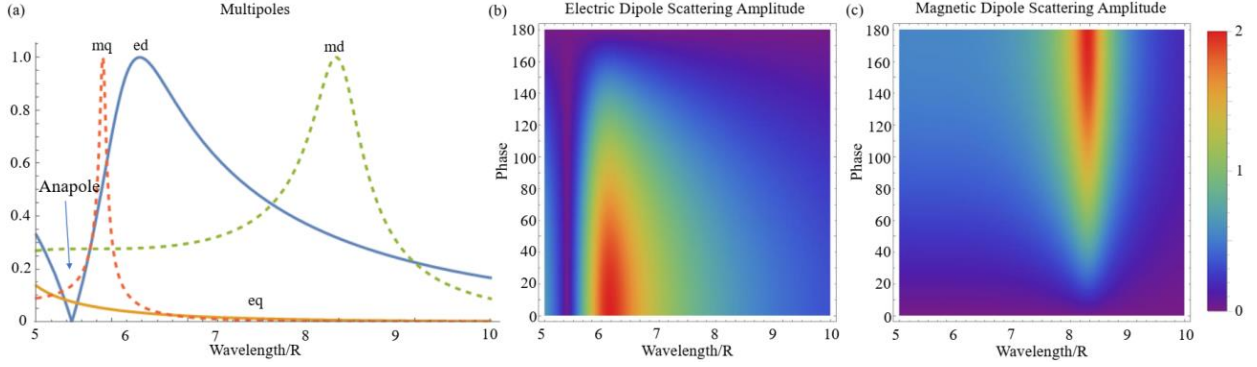

FIG. 7. Optical properties of the particle with refractive index n = 4. (a) Absolute electric and magnetic Mie scattering amplitudes on wavelength normalized to the radius ($\lambda/R$). (b) and (c) Absolute electric ($|a_1 + a_1 e^{i\varphi}|$) and magnetic ($|b_1 - b_1 e^{i\varphi}|$) Mie scattering amplitudes upon excitation by two plane waves, depending on $\lambda/R$ and phase difference of dipoles ($\varphi_d = |\varphi_1 - \varphi_2|$).

## Appendix B: Method of Coupled Dipole Equation

For a system of $N$ subwavelength interacting particles, the electric ($P_i$) and magnetic ($M_i$) moments of each particle can be found solving the following system of equations:

$$P_i = \alpha_i^e \left[ E_{id} + \sum_{j=1}^{N} \left( C_{ij} - \sqrt{\frac{\mu_0}{\varepsilon_0}} G_{ij} M_i \right) \right], \tag{B1}$$

$$M_i = \alpha_i^m \left[ H_{id} + \sum_{j=1}^{N} \left( C_{ij} M_j + \sqrt{\frac{\varepsilon_0}{\mu_0}} G_{ij} P_j \right) \right], \tag{B2}$$

where $\alpha_i^e$ and $\alpha_i^m$ are the electric and magnetic polarizabilities of $i$-th particle respectively

$$\alpha^e = \frac{3i\varepsilon_h}{2k_h^3} a_1, \quad \alpha^m = \frac{3i}{2k_h^3} b_1 \tag{B3}$$

where $k_h$ and $\varepsilon_h$ are the wavenumber and permittivity of the host medium, $a_1$ and $b_1$ are Mie scattering amplitudes. The quantities $E_{id} = C_{id} P_d$ and $H_{id} = \sqrt{\frac{\varepsilon_0}{\mu_0}} G_{id} P_d$ are the electric and magnetic fields of the elementary dipole source ($P_d$) respectively, in the point where the $i$th particle is located. $C_{ij}$ and $G_{ij}$ are the electric and magnetic fields of the point particle electric dipole of the $i$th particle at the $j$th particle with the following expressions

$$C_{ij} = A_{ij} I + B_{ij} (n_{ij} \otimes n_{ij}), \qquad G_{ij} = D_{ij} n_{ij} \times, \qquad i \neq j \tag{B4}$$

where $n_{ij} \otimes n_{ij}$ is the dyadic product of the unit vector of $r_{ij}$, which is the radius vector from $i$th particle to $j$th particle; $A_{ij}$, $B_{ij}$ and $D_{ij}$ are given by

$$A_{ij} = \frac{\exp(ik_h r_{ij})}{r_{ij}} \left( k_h^2 - \frac{1}{r_{ij}^2} + \frac{ik_h}{r_{ij}} \right) \tag{B5}$$

$$B_{ij} = \frac{\exp(ik_h r_{ij})}{r_{ij}} \left( -k_h^2 + \frac{3}{r_{ij}^2} - \frac{3ik_h}{r_{ij}} \right) \tag{B6}$$

$$D_{ij} = \frac{\exp(ik_h r_{ij})}{r_{ij}} \left( k_h^2 + \frac{ik_h}{r_{ij}} \right) \tag{B7}$$

Since only one particle (resonator) is used and an extra dipole is introduced in our paper, Eqs (B1) and (B2) have the form of

$$P_i = \alpha_i^e \left( E_{pd_1} e^{j\varphi_1} + E_{pd_2} e^{j\varphi_2} \right) \tag{B8}$$

$$M_i = \alpha_i^m \left( H_{pd_1} e^{j\varphi_1} + H_{pd_2} e^{j\varphi_2} \right) \tag{B9}$$

where subscript $p$ indicates the particle, $d_1$ and $d_2$ are first and second dipole with corresponding phase $\varphi_1$ and $\varphi_2$ respectively. After solving Eqs. (B8) and (B9), nonzero components of electric and magnetic dipole moments are obtained:

$$P_{py} = \alpha_p^e A_{pd} P_{dy} (e^{j\varphi_1} + e^{j\varphi_2}) \tag{B10}$$

$$M_{pz} = \alpha_p^m \sqrt{\frac{\varepsilon_0}{\mu_0}} D_{pd} P_{dy} (e^{j\varphi_1} - e^{j\varphi_2}) \tag{B10}$$

**Appendix C: Purcell effect in superdirective antenna**

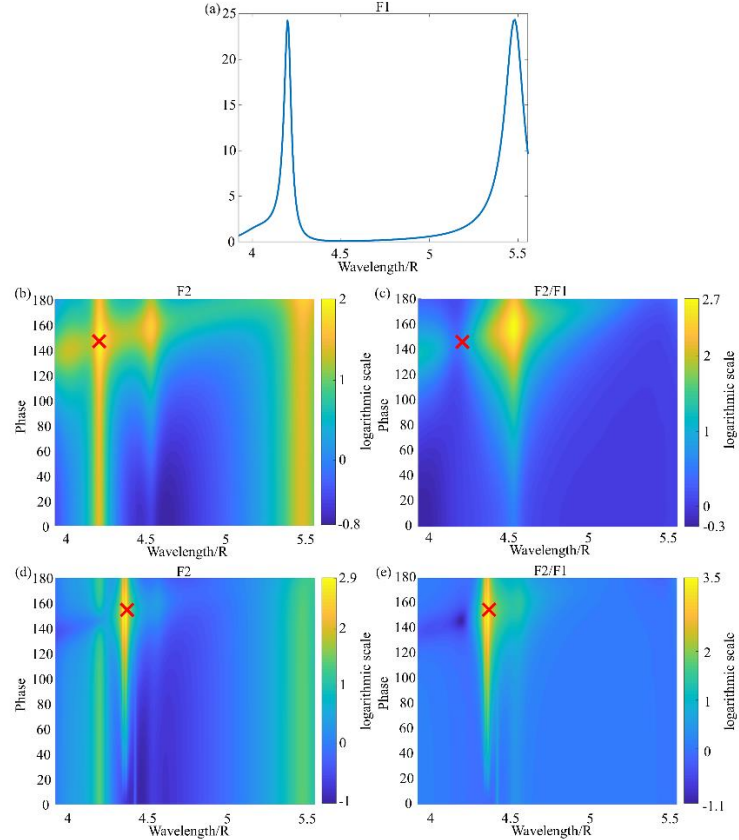

FIG. 8. (a) Purcell factor of a single dipole source in notched antenna, depending on $\lambda/R$. Collective radiation enhancement for two dipoles placing along the (b) x-axis and (d) z-axis in the notch. Normalized collective radiation enhancement for two dipoles placing along the (c) x-axis and (e) z-axis. The red crosses show the points of maximum total enhancement.